\documentclass[10pt]{article}

\usepackage{geometry}
\usepackage{float}
\usepackage{blindtext}
\usepackage{booktabs}
\usepackage{amsmath}
\usepackage{amssymb} 
\usepackage{enumitem}
\usepackage{epigraph}
\usepackage{lipsum}
\usepackage{authblk}
\usepackage[normalem]{ulem}
\usepackage[labelfont=bf, font=footnotesize]{caption}
\usepackage[numbers]{natbib}
\usepackage[dvipsnames]{xcolor}
\usepackage{lineno}
\usepackage{titlesec}
\usepackage{framed}
\usepackage{silence}
\makeatletter
\def\thanks#1{\protected@xdef\@thanks{\@thanks
            \protect\footnotetext{#1}}}
\makeatother

\titleformat*{\section}{\Large\bfseries}

\usepackage{graphicx}
\usepackage[colorlinks=true,linkcolor=Blue, urlcolor  = Blue, citecolor=Blue]{hyperref}

\usepackage{ragged2e}

\makeatletter
\renewenvironment{abstract}{%
    \if@twocolumn
      \section*{\abstractname}%
    \else
      \small
      \begin{center}%
        {\bfseries \abstractname\vspace{-.5em}\vspace{0pt}}%
      \end{center}%
    \fi}
    {% End of abstract environment
    }
\makeatother

\makeatletter
\renewcommand{\maketitle}{%
  \begin{center}
    {\LARGE \@title \par}
    \vskip 1em
    {\large \@author}
    \vskip 1em
    {\@date}
  \end{center}
  \vskip 1em
  {\def\thefootnote{}\@thanks} % Include the \@thanks content
}
\makeatother

 \newcommand{\response}[1]{#1} % When uncommented, this will print new text in black
 \newcommand{\remove}[1]{} % When uncommented, this won't print the old text

\title{\vspace{-.4in}
\Large \textbf{Capitalizing on a Crisis: A Computational Analysis of all Five Million British Firms During the Covid-19 Pandemic}\vspace{0.15in}
}

\author[1,2,3]{%
    \large{%
        Naomi Muggleton\thanks{%
            \noindent\textbf{For Correspondence:} Naomi Muggleton, Behavioural Science Group, Warwick Business School, Scarman Road, Coventry CV4 7AL, West Midlands. Tel: +44 (0)24 761 51327. Email: \href{mailto:Naomi.Muggleton@wbs.ac.uk}{Naomi.Muggleton@wbs.ac.uk}. \textbf{Acknowledgements}: This work was supported by funding from the British Academy (N.M., Grant number PF22/220090), the Leverhulme Trust for the Leverhulme Centre for Demographic Science (C.R., Grant number RC-2018-003), and the European Research Council (A.R., Grant number 849960). We acknowledge the use of the University of Oxford's Advanced Research Computing facility. We are especially grateful to Companies House for consultation and the provision of data. We would also like to thank Alexandra Rottenkolber for her help in preparing this manuscript, and Jason Bell and John Gathergood for providing feedback on a previous draft of this manuscript.
        }%
    }%
}

\author[4,5]{\large{Charles Rahal}}
\author[2,6]{\large{Aaron Reeves}}

\affil[1]{\small{Behavioural Science Group, Warwick Business School, UK}}
\affil[2]{\small{Department of Social Policy and Intervention, University of Oxford, UK}}
\affil[3]{\small{Brasenose College, University of Oxford, UK}}
\affil[4]{\small{Leverhulme Centre for Demographic Science, University of Oxford, UK}}
\affil[5]{\small{Nuffield College, University of Oxford, UK}}
\affil[6]{\small{Department of Sociology, London School of Economics, UK}}

\date{\today}
\begin{document}
\maketitle
%\vspace{-.2in}

\begin{center}
\textbf{Please cite this as:\\ \vspace{.1in}}
Muggleton, N., Rahal, C., \& Reeves, A. (2025). `Capitalizing on a Crisis: A Computational Analysis of all Five Million British Firms During the Covid-19 Pandemic', \textit{Journal of Computational Social Science}, \textbf{8}(2), pp. 1-29. DOI: \url{10.1007/s42001-025-00360-4}.
\end{center} \vspace{.1in}

\begin{abstract}
\small{
    The Covid-19 pandemic brought unprecedented changes to business ownership in the UK which affects a generation of entrepreneurs and their employees.
    Nonetheless, the impact remains poorly understood.
    This is \response{because}\remove{due to} research on capital accumulation \response{has} typically \remove{using aggregated data or small, non-representative samples,} lacked high-quality, individualized, population-level data.
    We overcome these barriers to examine who benefits from economic crises through a computationally orientated lens of firm creation.
    Leveraging a comprehensive cache of administrative data on every UK firm and all nine million people running them -- combined with probabilistic algorithms -- \remove{allows for}\response{we conduct} individual-level analyses to understand who became Covid entrepreneurs.
    Using these techniques, we \remove{identify detailed}\response{explore} characteristics of entrepreneurs \remove{including}\response{-- such as} age, gender, region, business experience, and \remove{sector}\response{industry -- which potentially predict Covid entrepreneurship}.
    By employing an automated time series model selection procedure to generate counterfactuals, we show that Covid entrepreneurs were typically aged 35-49 (40.4\%), men (73.1\%), and had previously held roles in existing firms (59.4\%).
    For most industries, growth was disproportionately concentrated around London.
    \response{It was therefore existing corporate elites who were most able to capitalize on the Covid crisis and not, as some hypothesized, young entrepreneurs who were setting up their first businesses. In this respect, the pandemic will likely impact future wealth inequalities.}\remove{The pandemic and ensuing government policies entrenched and exacerbated inequalities by age, gender, and region, pushing young people out of labor markets while creating opportunities for those already holding and controlling corporate capital.}
    Our work offers methodological guidance for future policymakers during economic crises and highlights the long-term consequences for capital and wealth inequality.
}
\end{abstract}\\


\newpage

\section{Introduction}

Who benefits from economic crises \citep{Walby2015}? One common hypothesis is that corporate elites are \response{differentially}\remove{more} able to capitalize on these moments of uncertainty, thereby expanding their wealth and power \citep{Carroll2008, Chowdhury2018,  Krippner2012}. However, the empirical evidence reveals a far more complicated picture \citep{Mizruchi2013}. When recessions occur, they typically involve the redistribution of income, resulting in both winners and losers \citep{Bargain2017, Cribb2017, Hellebrandt2014, Jenkins2012}. Interestingly, many of those who suffer economic losses are found at the top of the income distribution \citep{Jenkins2012, Piketty2013}. This redistribution of income can be sufficiently large that income inequality (as measured by \remove{-- for example -- }the Gini coefficient\response{, for example}) usually falls during recessions \citep{Cribb2017, Scheidel2017}. If economic crises can lead to declines in income inequality, then why do we assume that
corporate elites are able to exploit these moments of uncertainty to expand their wealth and power? Part of the reason is that these general declines in income inequality do not necessarily represent the fortunes of everyone at the top. Some business owners will see their personal fortunes decline during recessions because their income and wealth are heavily tied to their firm's financial performance \citep{RahmanKhan2012, Pfeffer2013, Sherman2017}. Conversely, other business owners will thrive amid the uncertainty, finding opportunities to generate excess profits by supplying new products or services, arbitraging prices, or experiencing an upturn in demand for their existing products or services, which become more valuable. The World Wars, for example, created significant capital destruction which contributed towards a general decline in inequality, while making arms manufacturers spectacularly wealthy \citep{Scheve2016}.

What is unclear, however, is the extent to which certain business owners \remove{(those who some have termed the `corporate elites') }are able to use these crises to expand their control over capital. We know that some business leaders have used crises to further their political agenda, such as through lobbying for self-interested policy changes \citep{Fransham2023, Hacker2011,Krippner2012}. We also know that, in general, many businesses close, and fewer businesses have historically opened during these economic crises \citep{Morris2009}. However, we know relatively little about whether recessions lead to a consolidation of economic power among some corporate elites, or whether recessions create \response{a form of `creative destruction' \citep{Schumpeter2013} which allows innovative} newcomers to start new businesses \remove{within a framework which may or may not be similar to Schumpeter's theory of `creative destruction' \citep{Schumpeter2013}}. We consider these questions by taking a `big', computationally orientated analysis of patterns in firm ownership during the widespread and dramatic decline in economic activity produced by Covid.
The dynamics of capital accumulation are pivotal to the production of wealth inequality, and so the concentration of wealth among a small group of firm owners not only \remove{distorts}\response{alters} the economic landscape but \response{can} also \response{amplify}\remove{amplifies} societal disparities. This small group of firm owners have sometimes been called the corporate elite, and they are typically those who have disproportionate access to and control over the resources of multiple firms \citep{RahmanKhan2012}, in part because they sit on the boards of more than one firm \citep{Larsen2018}. The enduring impact of capital on inequality broadly conceived -- underscored by the significant role of wealth accumulation in shaping economic power structures -- highlights the urgent need for a deeper exploration of ownership patterns \citep{Piketty2017, Saez2016}.

Recent studies utilizing Companies House data have largely focused on broad, aggregated metrics of economic activity during the Covid pandemic, such as the overall number of new business incorporations \citep{Duncan2020,ONS2021} or sector-specific effects (e.g., care homes \citep{Fotaki2023}, charitable organizations \citep{McDonnell2022}, Scottish universities \citep{Armstrong2023}).
Analysis of business creation during the UK lockdown highlights significant sectoral impacts, particularly in construction and retail \citep{Duncan2020}.
Similarly, Bahaj et al. \cite{Bahaj2024} examine the broader economic effects without analyzing individual-level characteristics of entrepreneurs.
Others have \remove{reflected this trend, providing }\response{provided }valuable insights into general economic disruptions and regional variations but fail\remove{ing}\response{ed} to capture the granularity of individual experiences (e.g., \cite{Fotaki2023, Wilson2023,Hurley2021}).
While these studies offer crucial context for the pandemic's economic repercussions, they overlook the realities of individual beneficiaries and losers during this time.
This research departs from these aggregate analyses by leveraging high-resolution, individual-level data to uncover detailed profiles of who profited amid the pandemic.
This view of individual outcomes across the entirety of incorporated companies is unrivaled in prior analysis of the corporate elite: it cannot be seen in aggregated data or self-reported survey data, and is only now possible through recent advances in computational approaches to public administration.

In an effort to explore whether some business owners, such as corporate elites, capitalize on crises we focus on firm creation and firm death during the pandemic and we do so because this particular recession limited the ability of the corporate elite to capitalize on the crisis in ways that were different from earlier recessions. This is because the spread of the disease -- combined with the associated policy responses -- created conditions that seemed to have been relatively favorable to new entrepreneurs. As noted above, during economic crises, the number of firms closing their doors increases and the number of firms being opened \remove{went}\response{goes} down \citep{Morris2009, Tian2018}. The economic downturn that occurred because of the Covid crisis was different \citep{Bartik2020, Humphries2020}. The pandemic appears to have substantially accelerated firm creation and slowed the number of firm closures \citep{Bartik2020, Baker2020}, partly due to several unique policies. Focusing on this particular recession is especially useful because if corporate elites were still able to capitalize on \textit{this} crisis then it is very likely that they would have done the same in earlier recessions which were less conducive to new entrants. This period of rapid firm creation\remove{, then,} helps us understand whether and how some corporate elites capitalize on economic crises. Indeed, the fact that the pandemic accelerated firm creation merely illuminates our `lack of understanding of how major macroeconomic events' affect the fortunes of corporate elites \citep{Santos2017}.

We therefore focus on one specific part of this puzzle: who were the beneficiaries of this economic moment \citep{Bahaj2024}? Two competing explanations exist for this acceleration in firm creation. The first is linked to the declining economic security of many people in the economy. Many workers -- despite government support \citep{Mayhew2020} -- lost their jobs or saw their incomes fall as businesses temporarily closed and people socially distanced \citep{Bell2020}. In this explanation, it is those who found themselves without work or on furlough, as suggested above, who used their newly found time to set up new businesses. Young people were hit especially hard by pandemic-related job losses, leading to speculation that these new firms might have been set up by members of this age group \citep{Wallace2021}, particularly as they are often well-equipped to use digital technologies (an increasingly necessary condition for firm creation). The second potential explanation focuses on existing corporate elites. As we've seen, many assume that this group is best placed to capitalize on the crisis in part because they had both the resources and the experience to create new firms. To address these questions, we collected administrative data on every single firm in the UK registered with Companies House. This represents approximately five million firms, and the near nine million people who ran them (between 2019 and 2021). This gives us an unparalleled perspective on who was creating these new firms.

We show that it was not young people and their `bedroom businesses' who benefited most from the crisis, but it was the `corporate elite', that is, those who already had \remove{strong links and }experience of running firms. As a result, the Covid crisis exacerbated age-based, gendered, and regional inequalities in firm ownership, potentially leading to a greater concentration of corporate control among those already holding positions, such as directorships, especially because firm survival is far higher among people who already own businesses. In doing so, our results contribute to broader debates about class stratification and the resilience of the capitalist class structure in times of economic crisis. Our analysis also speaks to debates \remove{beyond patterns of firm creation during the pandemic because it also shines a light }on the dynamics of capital distribution and firm ownership, revealing the broader trends of economic stratification and social mobility. These insights are crucial for understanding the evolving nature of wealth and power in the UK, and potentially offer a lens through which to view similar dynamics in other economies in the aftermath of major crises \citep{Beckert2022}. 

\subsection{Who are the corporate elite?}

\response{There is no single and widely accepted definition of the corporate elite. Some, following Marx, define the corporate elite as those who own capital while others, like Mills, view the corporate elite as those who control socially significant firms \citep{Scott1992}. Growing financialization has made the Marxist approach increasingly problematic because very few people are pure capitalists and many more people are partial capitalists in the sense of owning \textit{some} capital. The challenge with Mills' approach is that it is difficult to define the socially significant firms, especially because these firms change over time. While these approaches are certainly valuable, in this paper we draw on a related but different tradition which sees the corporate elite in terms of interlocking directorates or the ability to exercise control over a plurality of firms \citep{Bihagen2013, Carroll2008, Mizruchi2013}. Borrowing from this tradition, we conceptualize and operationalize the corporate elite}\remove{The corporate elite is often defined} as those who have significant control over more than one firm \citep{Carroll2008}.\remove{is often measured as}\remove{sitting on the board of more than one company}. \response{The advantage of this approach in our context is that our data gives us far more detailed information about the number of businesses these people are attached to than has previously been available, and we exploit that granularity in our analysis below.}

\response{Using the conceptualizations above, a number of scholars have tried to unpack the} \remove{The} social composition of this corporate elite\remove{ has long been a focal point within elite studies}, and \response{how this has}\remove{seems to have} changed over time. Historically, corporate elites -- like most other kinds of elites -- were disproportionately \remove{males}\response{men} drawn from affluent families who attended a small but prestigious set of schools and universities \citep{Rubinstein1981}.
For Mills \cite{Mills1956}, these corporate elites \response{joined together with political and military elites to form}\remove{formed} a cohesive group\remove{ with political and military elites} that was rooted in their shared trajectories through these socializing institutions. However, this is far less true today than it once was. There are more women in the corporate elite, and the presence of aristocrats has greatly declined \citep{Adams2023,  Keister2022, Reeves2024}. Similarly, although the power held by alumni of certain elite schools and universities (e.g., Oxford and Cambridge) diminished, it \remove{persists to some degree}\response{remains incredibly prominent} \citep{Reeves2017}. Importantly, some \remove{analyses}\response{accounts} focused not only on who elites were, but also on how their identities influenced \remove{both }their actions and the performance of \remove{these firms}\response{the firms they owned} \citep{LueckerathRovers2013, Post2015}. \response{One crucial dimension to these debates in Britain was the concern}\remove{Some of the early analyses in the British context were concerned} with the rapid decline of \remove{Britain}\response{it}'s economy compared to other high-income countries post-WWII \citep{Anderson1964, Miliband1969, Sampson1962}, a decline that was attributed to the gentlemanly culture that then dominated the economy \citep{Sampson1962}. \response{Critics of this gentlemanly approach to capitalism}\remove{Instead, they } called\response{, instead,} for a more dynamic, innovative, and, importantly, less class-ridden approach to Britain's leading businesses \citep{Adams2023}, and, as noted above, there is evidence that \remove{this} \response{the gentlemanly }nature of early twentieth century capitalism gave way to neoliberalism \citep{Miles2012}. 

We extend this tradition in two ways. First, we examine patterns of firm births and deaths among this corporate elite by using data on the entirety of firms registered in the UK ($\sim$5 million), plus the individuals who \remove{run (or have run)}\response{ran} them ($\sim$9 million). Secondly, through recent advances in data science\remove{and legislation that makes company ownership and involvement public}, we outline a methodology to identify detailed information about these individuals, including their gender, age, geographic location, industry of interest, and crucially, their experience in running firms. We are specifically concerned with whether certain kinds of corporate elites (those who already have significant control over existing firms) can capitalize on a crisis, and in this respect are interested in general patterns of who can create new firms (deepening their connection to the corporate elite) and who sees their firms cease trading (weakening their connection to that elite). We argue that these questions are \remove{sociologically}significant not only because they help us understand the degree of continuity within the capitalist class and the petite-bourgeoisie, but also because they allow us to understand the ways that corporate elites are (or are not) able to respond to moments of uncertainty. \remove{Do they capitalize on a crisis?}

\subsection{The \remove{sociological} roots of firm creation during Covid}

To understand the distributional effects of economic crises on firm owners, we need to understand why people start companies and how these reasons intersect with the particularities of the Covid crisis. The pandemic increased \remove{the}\response{people's} `intention[s] to start a new firm' \citep{Hill2022} \remove{and}\response{while} government support created broader and more favorable conditions for firm creation than previous economic crises. The `Job Retention Scheme' -- commonly known as `furlough' and which covered up to 80\% of people's salaries while they could not work -- may have fostered entrepreneurship precisely because people were still being paid despite a reduction in the hours spent on their primary form of employment \citep{Ebbinghaus2022, Mayhew2020}. In other words, if there was an economic crisis in which the concentration of ownership among the corporate elite was going to decline, it would be during Covid that we would see such declines. Early evidence suggested that some of this firm creation was in industries consistent with `bedroom businesses', such as online retail \citep{Bahaj2024}. But a rise in online businesses does not necessarily mean that these new firms are being started by people who are outside of the corporate elite. Moreover, there are good reasons to suspect that even though such outsiders experienced economic shocks, they would still have found it hard to start a business. Women in particular\remove{, for example,} disproportionately faced other care-related constraints that might have hindered their ability to become a `Covid entrepreneur'. Likewise, despite narrative anecdotes of the rise of bedroom businesses \citep{Wallace2021}, newly unemployed young people may have still faced significant financial barriers to establishing a new firm. Entrepreneurial activity takes both human and financial capital, and these are not evenly distributed throughout the population, potentially exacerbating inequalities in opportunity.

Such inequalities are crucial because new businesses are most often created by people who have experience of corporate administration, and who have support from social contacts and the institutional environment \citep{Zanakis2012}. An alternative hypothesis, then, is that existing business owners (\response{a subset of whom we regard}\remove{or what we might think of} as the corporate elite) were potentially uniquely placed to take advantage of this moment. They received extensive corporate welfare from the British government \citep{Farnsworth2013} and this support may have temporarily kept firm death to a minimum -- and even spawned new creations -- despite significant pressures from the government's decision to impose restrictions on everyday activities \citep{Newton2020}. Alongside this, many business owners likely already possessed the resources, experience, and professional networks needed to reduce frictions that pervade the process of firm creation \citep{Devece2016, Frese2023}. Indeed, although new firms in general have a low likelihood of succeeding, firms started by those with business experience (e.g., people in the corporate elite) are far more likely to survive and succeed. 

Two competing hypotheses \remove{have therefore emerged}\response{can be discerned in the existing literature}: `the rise in firm creation is driven by the newly unemployed, particularly by young people' and `the rise in firm creation is driven by existing corporate elites, specifically those with strong connections to existing firms'. \response{To be clear, we are less interested in the eventual success of these firms because we know that -- as suggested above -- firms established by experienced business owners are more likely to succeed. Rather, we are here primarily concerned with processes of firm creation. More precisely, the key debate motivating our analysis is whether the pandemic created conditions in which first-time business owners became more likely to establish new businesses. Moreover, we are not suggesting that these two hypotheses are mutually exclusive. Some new businesses were almost certainly set up by both of these groups. But we do see these hypotheses as being competing explanations because one of these explanations is likely to be the dominant reason for the rise in firm creation. Age is crucial here because public narratives emphasized young entrepreneurs (e.g., \citep{Wallace2021,BBC2022}) but, at the same time, it is also true that that young entrepreneurs are less likely to be persons of significant control in multiple firms compared to older entrepreneurs. In this respect then the key hypothesis we are testing is whether the rise in firm creation was driven more by first-time (likely to be young) entrepreneurs or corporate elites (likely to be older).} 

These two hypotheses \remove{matter }\response{have different implications for wealth inequality}\remove{because they potentially have profound implications for wealth inequality}. One primary reason for our focus on firm creation and demise is because firm ownership is a major (albeit not the only) source of wealth, and changes in firm ownership are important drivers of wealth inequality \citep{Beckert2022}. The UK's `Rich List', for example, is largely derived from information contained within Companies House \citep{Advani2022}, which is the primary data source that we use to address our hypotheses as described above. The salience of firm ownership to wealth inequalities can also be seen in studies which go beyond the `very rich'. Inequalities in firm ownership in the United States of America at the beginning of the 1960s are a major driver of the racial wealth gap still present today \citep{Lipton2022}. In Europe, income inequality is higher in countries where a larger proportion of firms are privately owned and where the ownership of publicly traded firms is more concentrated \citep{Peter2019}. On top of this, if firm ownership was shared more widely (i.e., through employee ownership schemes), then wealth inequalities would be dramatically reduced \citep{Dudley2021}. Thus, the surprising rise in firm creation during the pandemic could have important long-run implications for wealth inequality in the UK, \remove{rooted in the ownership and control of those firms. B}\response{b}ut the exact impact of this explosion of new firms depends in part on who started these business. Even though a large number of new firms end up closing within the first 5 years, if firm creation was dominated by the corporate elite (those who already ran businesses), then this could lead to even more entrenched patterns of wealth inequality in the UK in part because people who already run a business are far more likely to see their firms succeed than those who are starting their first business. But if this wave of firm creation was led by those who have traditionally been marginalized, then it could have an equalizing effect, albeit moderated by the number of closures among these new business.

Alongside the demographic characteristics of \remove{officers behind}\response{the people driving this} firm growth, there is also an important geo-spatial element. This matters because the UK is one of the most regionally unequal countries in Europe, with only Bulgaria being more unequal \citep{Dorling2022}. Regional inequalities in wealth are growing, which may also be a function of uneven developments in firm creation and death \citep{Fransham2023}. If firm growth is concentrated in London and the South East of England, then this too will likely exacerbate regional wealth inequalities over the coming years.

\subsection{Computational approaches to public administration}

Despite the UK leading much of the world in terms of `openness' in governance and public administration \citep{Brandusescu2018}, the depth of the literature that takes a computational approach to the UK's public sphere remains relatively shallow. This is despite the general rise in computational modeling across the social sciences \citep{Rahal2022a}, and the widespread, high-profile use of administrative data for research in the United States (see, for example, \cite{Bell2019}). There is, however, an emerging body of work which is proximal to our own in a UK context. Clifford \cite{Clifford2018}, for example, uses longitudinal registry data on 125,000 charitable organizations in the UK to analyze the density of charitable organizations. Rahal and Mohan \cite{Rahal2022b} use both this same Charity Commission data as well as information from Companies House (as done in this paper) to `reconcile' millions of payments made across the entire hierarchy of the National Health Service through a computational approach which involves custom ElasticSearching and \texttt{.pdf} parsing. In the most similar computational analysis of Companies House to our own, Bahaj et al. \cite{Bahaj2024} studies newly created firms by merging market entry data with online job postings to show that increases in firm creation drive increased vacancy postings.

\response{We draw on these computational approaches to administrative data to address significant gaps in our understanding of who can capitalize on a crisis, using the Covid pandemic as a case study}\remove{Currently, we know very little about who benefited from the Covid pandemic}. \response{In other words, we ask: d}\remove{D}id this period of uncertainty lead to a concentration of corporate power, or provide space for new entrepreneurs? \remove{We fill this gap through a computational re-examination of three rich, complementary datasets.}

\section{Data}

\subsection{Companies House data}

\remove{The first and most important data set that underpins our analysis is}\response{Our research design is underpinned by} administrative data from Companies House (the corporate registrar of the UK).
\response{We utilize the Free Company Data Product for information on active firms (where differences between monthly files indicate change of status; July 2019-June 2021), and monthly company officer snapshots (February 2019-June 2021). Companies House collects data and makes it publicly available under the Companies Act 2006. With the exception of a small category of material which is exempt from statutory disclosure requirements, Companies House is required by law to make the information available for public inspection. Information on the public register is made available by virtue of approvals issued in accordance with Section 47 of the Copyright, Designs and Patents Act 1988 and Schedule 1 of the Database Regulations (SI 1997/3032). Companies House imposes no rules or requirements on how the information on the public register is used. This administrative dataset includes records of approximately five million firms and nearly nine million individuals who have held officer positions from January 2019 to December 2021. The data has both cross-sectional and longitudinal components, allowing us to track changes over time, such as firm openings and closures, and officer appointments and resignations.
We accessed the data through Companies House's bulk data download service, which provides comprehensive and up-to-date records.}

\subsubsection{\response{Officer data}}

\response{Data on company officers was sourced from the Companies House FTP server \citep{CompaniesHouse2021} between July 2019 and June 2021. The data are based on monthly snapshots of company status, plus a list of all officers appointments and removals.}
\remove{This data provides information on those in positions of `significant control' of UK companies, who the `officers' of the companies are, and meta-data on the firms themselves. All new firms are required to register with Companies House (as a function of the Companies House Act of 2006) and -- except for a small category of material which is exempt from statutory disclosure requirements -- Companies House is then required by law to make the information available for public inspection. }\response{The officer data includes the following components: officer identifiers, names, dates of birth, appointment and resignation dates, associated firm identifiers.} \remove{This extraordinarily high-quality and granular data contains standardized information that allows us to engineer firm births and deaths on a monthly basis, alongside data on persons of significant control and officer-ship which includes names, ages, nationalities, and registered addresses. }\response{The data has a longitudinal aspect, enabling us to observe changes over time, such as when firms open or close, and when officers are appointed or resign.} We focus our analysis on firm officers, as opposed to firm owners\remove{.}\response{, as i}\remove{I}t is regularly the case that officers are in fact owners in our data: most firms are small- to medium-sized enterprises and in these firms, the listed officers (i.e., the directors) are also `persons of significant control' (i.e., owners). In large firms -- where the `persons of significant control' are \remove{usually}\response{sometimes} other companies -- officers are commonly shared across both the parent company and the subsidiary. Analyzing officers then provides us with a rich picture of the range of people who have an active stake in these firms.

\subsubsection{\response{Company metadata}}

\response{Metadata on companies was sourced from the `Free Company Data Product' of Companies House \cite{CompaniesHouse2021}. A monthly snapshot of all companies registered in the UK was scraped for all months between July 2019 and June 2021 inclusive.}
\response{The company metadata includes the following components: firm identifiers, registration and dissolution dates, statuses, Standard Industrial Classification (SIC) codes, registered addresses.}

\subsubsection{\response{Historic data}}

\response{The Companies House data is extraordinarily high-quality and granular, providing standardised information on firm births and deaths on a monthly basis, as well as data on persons of significant control and officers, including their names, ages, nationalities, and registered addresses. Despite its strengths, one limitation of Companies House data is the lack of company outcomes data prior to July 2019. This constrains our ability to construct long-run time-series forecasts, which typically require an extended historical baseline for improved accuracy. To address this limitation, we integrate two supplementary datasets. The first is the Fame dataset \cite{FAME2020}, which extends the temporal coverage of company outcomes such as openings and closures by up to 20 years. Using this data, we construct robust time-series forecasts for firm counts, relying on Fame data from January 2011 to January 2020 to complement the Companies House records. To identify trends in the number of companies registered prior to 2012, we utilized Companies House data on the total number of companies active in the UK between 1939 and 2022.\cite{CompaniesRegister2021}}

\subsubsection{\response{Data preprocessing}}

\response{We utilized computational methods to process and analyze this extensive dataset.
The large scale of the data necessitated efficient data processing and management techniques.
We developed scripts in Python to automate data cleaning, merging, and matching processes, enabling efficient handling of millions of records.}

\paragraph{\response{Officer preprocessing}}

\response{Data on company officers was sourced from the Companies House FTP server \cite{CompaniesHouse2021} between July 2019 and June 2021. The data are based on monthly snapshots of company status, plus a list of all officers appointments and removals, where differences between monthly files indicate change of status; July 2019-June 2021. Officers that were other companies (rather than people) were identified by Companies House's classification system, plus regular expression pattern detection for the following words in the Officer's Name:}

\begin{quote}
    \response{`COMMERCIAL', `COMPANY', `CORPORATE', `DETAILS', `EXCHANGE', `HOLDINGS', `INTERNATIONAL', `INVESTMENTS', `LIMITED', `LTD', `NON-DESTRUCTIVE', `PARTNERSHIPS', `PRIVATE', `PROSECUTION', `SECRETARIAT', `SERVICES'.}
\end{quote}
    
\response{We estimate that 3.42\% of officer appointments and registrations are those held by other companies. These were subsequently removed from our analysis.}

% \response{To be more precise, our corporate data wrangling and individual reconciliation approach allows us to create time series variables and projections as well as cross-sectional data which can be linked to geographical shapefiles and concatenated into specific industries. It also allows us to create aggregate metrics across age and gender.}\remove{In the majority of our analysis, we simply consider the change of a metric of interest across two points in time for a cross-sectional unit. It also provides detailed (and self-reported) Standard Industrial Classification (SIC) codes of economic activity which specify the industry in which these firms operate. We map these into twenty-one categories based on the first two digits of the code.}

\paragraph{Gender and age}

\response{We inferred missing demographic attributes such as gender and age from the available data fields, which is essential for our analysis of the socio-demographic composition of firm officers.}
Gender \response{may influence who creates new firms and has been central to debates about the corporate elite more broadly}\remove{is a key intersectionally stratified component of who creates firms, and in debates around the corporate elite more broadly} \cite{worth2024almost}. The gender of officers is not directly reported by Companies House, and so we inferred \response{gender was inferred using the \textsf{Python} packages \texttt{gender\_detector} \cite{Vanetta2015} and \texttt{gender\_guesser} \cite{Perez2016}, where residuals were sent as calls to an API service (\url{genderize.io}, as described in \citealp{Ehrhardt2018}).
These inferences involve probabilistic algorithms and machine learning estimators.
    To maximize the accuracy of gender estimates, we specified the appropriate locality as the United Kingdom or Great Britain for each API.
    In the event of disagreement between APIs, we opted for the most common estimate.
    Using this estimate, 1.6\% of officers could not be reliably labelled as men or women, and were subsequently excluded from all gender analysis.
    Of the remainder, 28.9\% were labeled as women and 71.1\% were labeled as men.}\remove{gender using commonly available tools (such as but not limited to that of, \cite{Perez2016}), specifying that these names were based in the UK or Great Britain.
In some cases, there was disagreement between classifiers, and in these cases, we opted for the most common estimate. Using this approach, 1.6\% of officers could not be reliably inferred as male or female and were subsequently excluded from all gendered analysis. Of the remainder, 28.9\% were classified as women, and 71.1\% were classified as men.} We acknowledge that such tools have limitations, not least the fact that algorithmic predictions reproduce binary notions of gender and even erase some genders \citep{Lockhart2023}. We have confidence in the inference of gender given that our estimates are similar to others drawn from self-reported data in samples \citep{Buchanan2023}\remove{.}, and briefly discuss potential measurement error which arises in approaches such as this in Section \ref{sec:discussion}. The month and year of birth of the officers is reported by Companies House data, and so a deduction of age is possible by subtracting an individual's month and year of birth from the month of observation. \response{For simplicity, all individuals were assumed to have been born on the first of the month. For example, if John Smith was born in January 1989, he would be labeled as 30 in observation month January 2019 and 31 in observation month January 2020.}

\paragraph{Location}

We inferred officer's location using the `Correspondence address' field within Companies House data, converting postcodes into Lower Layer Super Output Areas (LSOA) using \url{postcodes.io}. Regional labels were derived from the prefix of postcodes. Postcodes listed as being outside of the UK were excluded from the analysis. This is a slightly imperfect measure of the geographical location of company officers, and we discuss the possible implications of this in later sections below.

\paragraph{First-time officers and the corporate elite}

\remove{We operationalize} \response{Our primary operationalization of} the `corporate elite' \remove{as} \response{is} those who \response{start a new firm given that they} are already officers of at least one firm (and more commonly than not, individuals \response{are }hold\response{ing} multiple officer-ships). This means that we need to identify whether an individual was a `first-time' officer, or whether they had acted as an officer in a company previously. This does not mean that we think of every business owner as a member of the corporate elite. Rather, they become part of the corporate elite once they have started a second business or become an officer in a second business. We performed entity resolution to determine whether an individual was a first-time officer or had prior appointments. This involved matching individuals across different records using combinations of full names and dates of birth. Recognizing that names might have variations (e.g., use of middle names, initials), we employed fuzzy matching techniques using the Levenshtein distance algorithm. A subset of matches was manually checked to assess the accuracy of the automated matching process. To do so we grouped observations officer $\times$ company observations ($N$ = 20.3 million) for all officers who had been active at any firm since the introduction of the Companies Act in 2006.

\response{As noted above, there is no accepted definition of the `corporate elite' and even Mills' classic text \cite{mills_power_elite_1956} fails to provide a clearly delineated account of who is and who is not part of this group. We focus on those who are persons of significant control in at least two firms for a few reasons. The first is that the classic literature on `interlocking directorates' uses a very similar approach to defining the corporate elite \citep{Bihagen2013, Carroll2008, Mizruchi2013}, and does so because this captures people with a relatively high degree of corporate power. The second is that people who own at least two firms in Companies House are very likely to be in the top 10\% of the wealth distribution and maybe even be in the top 1\%, depending on how successful those businesses were. One of the notable differences between the top 10\% of the wealth distribution and the decile just below them is the likelihood that they own a business. Around 12\% of the wealth held by the top 10\% is linked to business wealth while it is only around 1\% in the decile just below that. Third, the proportion of people attached to two or more firms is only 5\% of the adult population (aged 18+), suggesting that our definition is capturing a small but wealth group of people.}

To measure this, we used the data from the population of officers contained in Companies House, and calculated the number of companies each officer had been attached to since the introduction of the Companies Act in 2006. We disambiguated officers according to whether the first name, surname, and month and year of birth were identical, and then counted the number firms to which they were attached. If the count was greater than one, then we can reliably infer that they had previously been an officer in a now deceased firm, or in a firm that was active at the time.

\response{It is important to recognize who is included in this definition of the `corporate elite'. Imagine someone opened their third small corner shop during the pandemic. Would they be part of the corporate elite? Probably not. This is because these stores would likely be registered under one corporate entity. However, if this person with three stores registered under one business then started a new business during Covid, in addition to the grocery stores they already owned, then we would classify them as being part of the corporate elite \citep{Advani2020}. We recognize, however, that this approach to operationalizing the `corporate elite' may be too inclusive and our results could therefore be contested on these grounds, and so we replicate our analysis using a variety of different thresholds. We start, as mentioned above, with people who establish a second firm during Covid.
Next, we look at people who started a third firm, then people who started a fourth firm, and so on. This allows us to observe whether the basic patterns in our analysis are consistent even when using more restrictive thresholds to define the corporate elite.}

\paragraph{Changes in firm status}

Data on when companies opened or closed was sourced from the `Free Company Data Product' of Companies House. We pro-actively collected monthly snapshots of all companies registered in the UK for all months between July 2019 and June 2021 inclusive during those months. In each month of observation, we automatically classified companies as `Opened', `Closed', `Reopened', or `No change'. In each month (`$t$'), firms would be labeled as `Opened' if they appeared on the register of companies in time $t$ for the first time (i.e., was not present at $t-1$ or earlier). Firms would be labeled as `Closed' if they were labeled as one of the following in period $t$: `Active - Proposal to Strike Off', `Administration Order', `Administrative Receiver', `In Administration', `In Administration/Administrative Receiver', `In Administration/Receiver Manger', `Receiver Manager/Administrative Receiver', `Voluntary Arrangement/Receiver Manager'. This definition was approved in private conversations with representatives at Companies House. Firms would be labeled as `Reopened' if the company was labeled as `Active' in the past, but `Closed' in period $t-1$ (above) and `Open' in $t$. Finally, companies are denominated as `No change' if the company was labeled as `Active' in both $t-1$ and $t$. \response{This computational approach to data wrangling and officer reconciliation allows us to make time series and cross-sectional analyses (across strata such as gender, industry, and geography) in order to analyze our key research question.} We also collected historical data from Companies House (`Companies Register Activities') on the number of companies registered in each year in the UK between 1939 and 2022, allowing us to look at long-term trends in the balance between firm creation and firm closure.

%\subsection{\remove{Supplementary data}}

\remove{Alongside this administrative data, we also examine national-level employment rates between 2011 and 2021 (from the Labour Force Survey) and self-reports of employment outcomes via a large longitudinal household panel dataset for a representative sample of the UK population between 2009 and 2021 (Understanding Society). National level data is used to calculate employment-population ratios by dividing employment by the total population, where total population is the sum of the employed (Dataset X01: Labour Force Survey single-month estimates), unemployed (X02), and economically inactive (X03). Understanding Society data were collected monthly between April to July 2020 and then every two months until June 2021 for over 11,000 households, with Covid modules grounded in the properties of the original data collection procedures and defined to provide reliable population inferences.}

\section{Methods for counterfactual creation}

\response{Our primary research question is whether the Covid crisis precipitated a change in the composition of those creating new firms. A full population of newly registered firms allows us to conduct deep descriptive analyses of the composition of firm \response{officers}\remove{owners (and other persons of significant control)} over time. In addition, to these deep descriptive analyses, and to}\remove{To} compare changes at the firm and individual-level, we estimate the economic impact of the Covid recession relative to estimated projections of what would have occurred in the absence of the pandemic. \response{This approach allows us to create a relative counterfactual; for each time series, what would have happened in the absence of the pandemic? This projection-based approach is necessary, because more traditional tools of causal inference are rendered obsolete as all strata of the UK economy were `treated'.} Our primary dependent variable in this analysis is the net change in the number of active firms, defined as the difference between the number of firms created and the number of firms that ceased operations within each month of our study period. This measure -- which we term `excess economic loss' -- is analogous to the approaches used to calculate excess mortality and provides a direct measure of the pandemic's impact on the entrepreneurial landscape of the UK \citep{Polyakova2020}. In this approach, economic loss is defined as the difference between a projected value ($\hat{y}_{t+n}$) of our dependent index at $n$ periods into the future, and its realized value ($y_{t+n}$). We follow the \texttt{autoarima} approach of \cite{Hyndman2008} by using a stepwise procedure to estimating the optimal Seasonal AutoRegressive Integrated Moving Average (SARIMA) (max $p$=12, $d$=1, $q$=12, $P$=12, $D$=1, $Q$=12). Optimal models were selected on the basis of minimizing the AICc, which is a a second order correction to AIC \citep{Akaike1998} which attempts to further minimize overfitting. The SARIMA($p$,$d$,$q$,$P$,$D$,$Q$) process is given by:

\begin{equation}
    \Phi(B^m) \phi(B)(1-B^m)^D(1-B)^d y_t = c +\Theta (B^m) \theta(B) \varepsilon_t
\end{equation}

\noindent{where} $\Phi$(z) and $\Theta$(z) are polynomials of orders $P$ and $Q$ respectively, each containing no roots inside the unit circle, and $\varepsilon_t$ is a white noise process with mean zero and variance $\sigma^2$. $B$ is a backshift operator, and if c $\neq$ 0, there is an implied polynomial of $d$ + $D$ in the forecast function. $m$ is the seasonal frequency of the data being modeled. Unit root tests are used to calculate the necessary level of differencing in the form of the KPSS unit-root test \citep{Kwiatkowski1992}: a test which has a null hypothesis that an observable time series is stationary around a deterministic trend. 
 Seasonal differencing is undertaken by a \response{Canova and Hansen} \cite{Canova1995} test: a test statistic \remove{for}\response{of} the null hypothesis that the seasonal pattern is stable.\footnote{\response{While it is not our primary methodological tool, we also conduct a number of tests for structural breaks as reported in Section \ref{sec:results}. The methods employed include: a CUSUM test \citep{page1954continuous}, a Chow test \citep{chow1960test}, a Bai-Perron test \citep{bai1998estimating}, Pettitt's Test \cite{pettitt1979nonparametric}, and a Zivot-Andrews test \citep{zivot1992structural}.}} All statistical tests conducted within this paper are two-sided. The value of this approach is that it allows us to consider whether specific socio-demographic groups were more negatively impacted than others, and to compare the direction and magnitude of our findings with the effects of the `Great Recession' of 2008-09. Our Covid-specific window for which we create counterfactual analysis (i.e., $\hat{y}_{t+n}$\remove{, }) spans from March 2020 to June 2021. The beginning of this time frame signifies the month in which Covid-related economic and social policies were announced, the ensuing Covid recession in the UK (Q1-Q2, 2020), and the subsequent twelve months of recovery following the recession.

\section{Results}\label{sec:results}

\subsection{Is firm growth driven by declining firm death or rising creation?}

\begin{figure}[!t]
    \begin{center}
        \includegraphics[width=\textwidth]{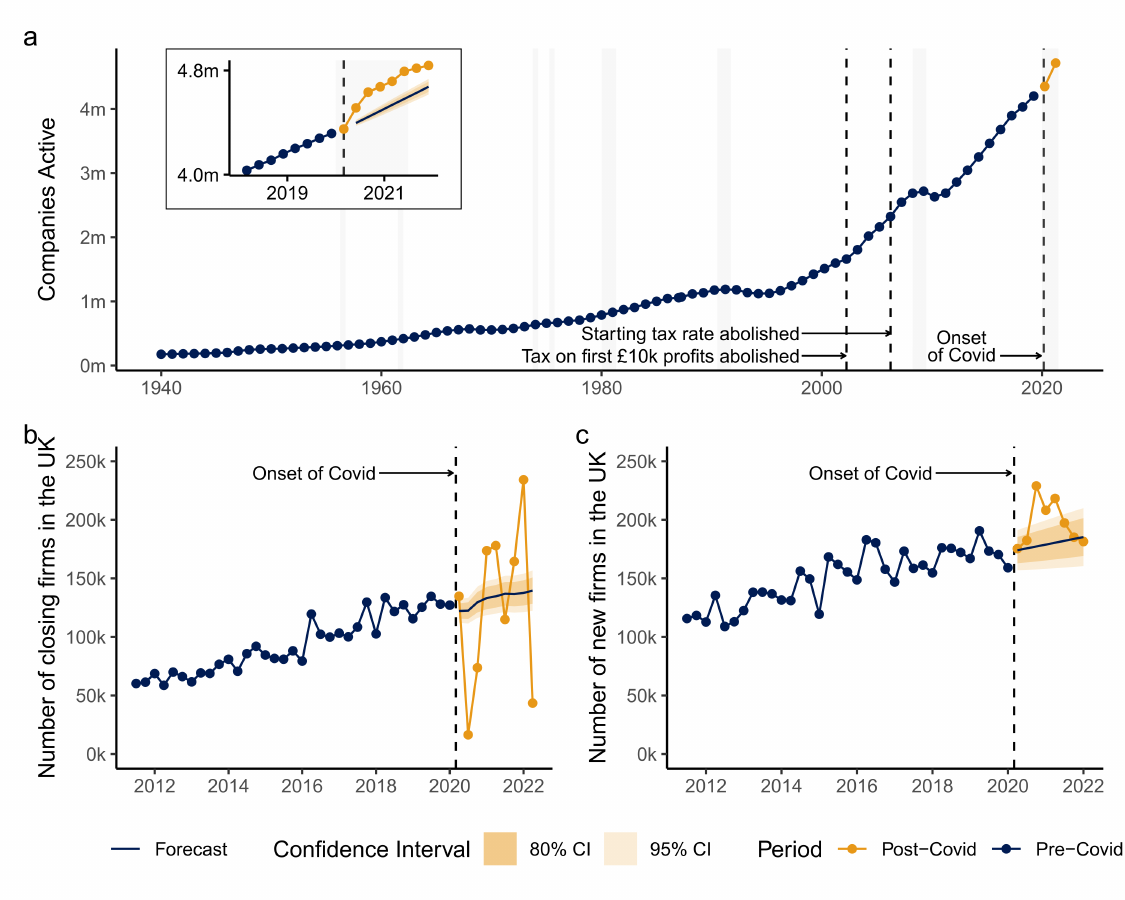}
        \caption{
        \response{Top panel: number of companies registered in the UK. Main plot shows annual data, with vertical black dashed lines highlighting administrative change at Companies House and shaded grey bars denoting periods of recession in the UK. Inset shows quarterly data from 2018 to 2022, with shading denoting 95\% and 80\% confident intervals, where the gold line denotes observed number of companies registered in the UK, and the navy line denotes our projections based on observations between January 2011 and January 2020. Bottom panel: n}\remove{N}umber of firms opening (\response{b}\remove{left panel)}) and closing (\response{c}\remove{right panel}) as on Companies House at each quarter between Q1-2011 and Q1-2022. Navy points denote observed time-series data made available by Companies House, and the solid navy lines denote output from our seasonally adjusted ARIMA models trained on data between January 2011 and January 2020 (the last quarter prior to the Covid recession). Yellow line denotes observed out-of-sample forecast data that was not used in the SARIMA models, along with 80\% and 95\% confidence intervals.}
        \label{fig:ch_time}
    \end{center}
\end{figure}

We start by outlining the trends in both firm death and firm creation before unpacking whether the corporate elite or new Covid entrepreneurs were driving macro-level trends.\remove{Figure~1 shows the impact of the crisis on employees and employers across the whole of the UK.}\remove{The story pertaining to the labor market is well-known. There was a sustained decline in the national employment-to-population ratio after March 2020 that persisted until March 2021 (Figure~1a.-1b.). At the same time, household incomes stagnated during the second and third quarters of 2020 (Figure~1c.), with income growth resuming in the fourth quarter.}\remove{This} \response{Covid} was a major economic shock, one that had the potential to harm the prospects of business owners. However, contrary to expectations, the number of active companies grew substantially over this period: even faster than projected\remove{ (Figure~1e.)}\response{.} In July 2021, there were nearly 180,000 (179,108, 95\% CIs [65,044, 293,172]) `excess' firms ($\hat{y}_t - y$) operating in the UK (Figure~\ref{fig:ch_time}a). This higher-than-expected number of firms was attributable to both fewer firms closing, and more firms opening (Figure \ref{fig:ch_time}b, c). 

\noindent{\footnotesize{\remove{\textbf{Figure 1:} The economic impact of the pandemic. Vertical gray fill in time series sub-figures denote recessionary periods in the UK (a-b and d). Quarterly employment rates for men (a) and women (b) in the UK between January 2010 and June 2021. Yellow and blue lines denote macro-level time-series data (ONS), and black lines denote out-of-sample forecasts from our models trained on data from between January 2010 and January 2020. Sub-Figure c: Monthly gross income household income. Sub-Figure d: Monthly gross income for males and females (with data for c and d from Understanding Society, with data for c coming from Covid waves). Sub-Figure e: Number of companies registered in the UK. Main plot shows annual data, with vertical gray dashed lines highlighting administrative change at Companies House. Inset shows quarterly data from 2018 to 2022, with shading denoting 95\% confident intervals, where the solid line denotes observed number of companies registered in the UK, and the dashed denotes our projections based on observations between January 2011 and January 2020.}}}

In the first quarter of the Covid recession, firms closing fell to just one seventh of our synthetic\response{, SARIMA-based} counterfactuals, meaning that around 100,000 firms (105,969, 95\% CIs [94,961, 116,977]) continued to operate in mid-2021 that would have likely closed were it not for the pandemic (Figure~\ref{fig:ch_time}b). \remove{Despite the fact that there was a moratorium (as a function of The Corporate Insolvency and Governance Act, 2020) on firms closing during the initial months of the Covid recession, there is no significant evidence to suggest that the majority of the firms that would have closed in this period then went on to do so after the provisions ended. }\response{This reduction in closures was primarily driven by the moratorium on firm closures introduced under The Corporate Insolvency and Governance Act, 2020 \cite{keddie2020corporate}. Following the end of this moratorium, the number of firm closures exhibited substantial fluctuations, likely due to bureaucratic adjustments processing the backlog. Some months saw a surge in closures, while others saw almost none. Given the administrative nature of these fluctuations, our analysis focuses instead on firm creation, which provides a clearer and more consistent indicator of economic dynamics during the pandemic. We also ran structural breaks on six series; every variation of seasonally and non-seasonally adjusted, and the time series for first `opened', `closed', and of `no change'. Many tests indicated a change at either 2020-03, 2020-04, or 2020-09. Further details are available as part of our replication materials.}

Crucially, the rise in the total number of firms was not just driven by reductions in closures; the number of firms being created rose by 25.2\% in the first quarter after the pandemic (Figure~\ref{fig:ch_time}c). This is a remarkable change in a relatively short period of time, and we estimate that 51,716 (95\% CIs [32,108, 71,324]) additional `excess' firms were created in this quarter alone. It is this increase which forms the heart of the paradox \response{we discussed in the introduction}. On the one hand, firms were jettisoning workers in large numbers, while on the other hand, the number of firms was growing. At a national level, the first way that the pandemic benefited \remove{the }business owners in general and the corporate elite more specifically was by keeping their operations going longer than expected. Indeed, \response{these `expected' firm deaths had not materialized by the end of our study period.}\remove{ what is striking about this is that by the end of our study period, expected firm deaths had not materialized.} This means that the pandemic saved \remove{-- at least temporarily -- }a non-trivial number of firms from closing and allowed them to operate for as much as twenty months longer than expected. Many of those firms that we would have expected to close in the absence of the pandemic were likely to be companies that were either struggling to remain profitable, or which had slipped into inactivity. The pandemic then, in many cases, either made it financially viable to keep those firms active or enabled them to become financially viable once again in operational terms.

The most obvious short-term explanation for this was almost certainly the financial help provided by the government in the spring of 2020. Many of these firms that were close to formally shutting their doors were propped up by the `Business Grants', `Bounce Back Loans', and the `Eat Out to Help Out' schemes. But these forms of support should not necessarily have made these business financially viable over the medium-term; that is, we still would have expected many of these firms to close in 2021. We see no acceleration of firm death through the second year of the pandemic, which suggests that something more fundamental has happened to keep many of these firms active. The first way then that the pandemic benefited business owners in general and the corporate elite more specifically was through a form of corporate welfare \citep{Farnsworth2013} which ensured that firms on the brink of \response{c}losing stayed open much longer than expected, and indeed were yet to close by the end of 2021.

\subsection{Which groups drove firm growth?}

\begin{figure}[H]
\begin{center}
    \includegraphics[width=\textwidth]{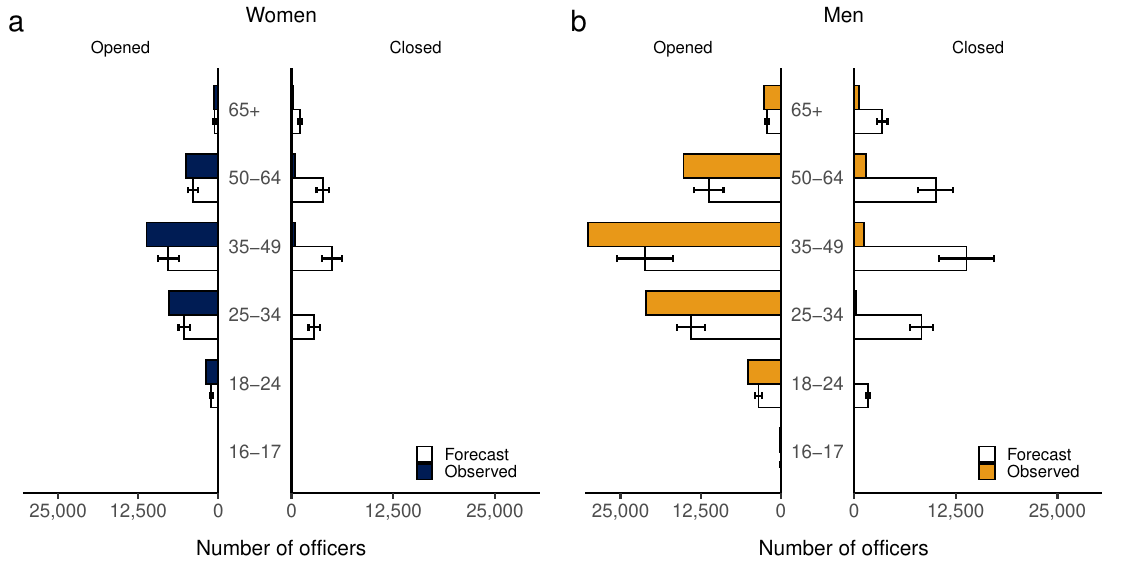}
    \captionsetup{width=\textwidth}
        \caption{No significant gendered difference across male and female officers. The bar and error plots show in both sub-figures and that there was no significant gendered difference between male or females other than the absolute number of officers across opening and closing firms. There was, however, a stratified difference across age groups.}
        \label{fig:double_bars}
    \end{center}
\end{figure}

If the pandemic increased the total number of firms, then which groups benefited from this growth in new businesses? First, we look at gender, comparing the experiences of men and women in the labor market with the experience of new business owners. We do this because the economic loss faced by employees and employers was stratified by gender. The UK's first major lockdown closed schools and offices \citep{Mayhew2020}, reducing employment rates, working hours, and earnings for men and women\response{, although the labor market effects of the Covid recession were -- in relative terms -- unusually gendered}. \remove{But men seemed to fare worse. Aggregate macroeconomic data on employment from the Office for National Statistics shows us how -- in June 2021 -- the employment rate for men was 2.6 absolute percentage points (95\% CIs [1.4\%, 3.7\%]) lower than projected. For women, the employment rate in the same period was 1.9 absolute percentage points (95\% CIs [0.9\%, 2.8\%]) lower. In other words, at the aggregate level, a larger proportion of men had lost their jobs than women. Turning to the micro-level, we see a similar pattern: for those who remained employed throughout the pandemic, pay fell by 13.7\% (\pounds340) for men, and 8.2\% (\pounds141) for women. Working hours too fell slightly more for men (29.9\%, or 11.2 hours) than for women (27.7\%, or 8.2 hours) too. When viewed in absolute terms, men seemed to fare worse than women in this pandemic induced recession. But this perspective shifts when viewed in the context of the Great Recession of 2007-2008, where male employment fell by considerably more than female employment.} \remove{The labor market effects of the Covid recession were -- in relative terms -- unusually gendered.}Yet the impact of the pandemic on firm owners and officers was largely gender neutral. Women are, of course, less often firm owners than men \response{in absolute terms} (Figure \ref{fig:double_bars}, Table \ref{table:sic}), but the pandemic has done little to close this gendered gap. 

Next, we consider the age dimension. Labor market shocks affected the youngest and oldest in society most acutely \response{while those in the middle of their working lives did not see significant labor market losses in terms of unemployment or income.} \remove{(Figure \ref{fig:double_bars}). Women aged 16-17 suffered a loss of 2.1\%, whereas for men aged 18-24, employment loss was 3.1\%. Additionally, men and women aged 50 and over suffered significantly lower employment numbers than projected. In contrast, those in the middle of their working lives -- women aged 18-49 and men aged 25-64 -- did not see significant labor market losses in terms of unemployment.} The effect of the pandemic on firm officers, however, was similarly felt across all age groups, although in different absolute numbers. Prior to the pandemic, the mean age of firm officers ranged from late 40s to early 50s across all industry sectors (Table \ref{table:sic}). To test whether this composition changed in light of the pandemic, we compare the number of firms opening and closing in the quarter preceding July 2021 (the first full quarter following the onset of Covid) to pre-pandemic estimates (SARIMA model in-sample range: January 2005 to January 2020). There were significant changes -- \remove{when }compared to pre-pandemic projections -- in the number of firms opening and \remove{the number of firms }closing \remove{(one quarter after the onset of the pandemic) }for all age groups (aged 16 to 65+).

\response{Yet, the size and direction of these changes varied by age.}\remove{There was, however, a difference in how much firm creation and firm death there was among different age groups.} For example, firm closures \textit{increased} for those aged 18-24 (vs. a pre-Covid baseline) by 1.3\%, but \remove{firm closures }fell for all \response{other }age groups \remove{over 25}. The biggest decline in firm death was experienced by officers in middle-aged groups (7.6\%)\remove{; for over 65s, the decline in firm death was just 3.7\% relative to pre-Covid baselines}. \response{Young people did see a rise in firm creation but this was dwarfed by a much larger rise in firm creation among the middle-aged}\remove{In terms of firm creation, the growth rate was highest amongst those in the 18-24 age range (36.4\%), and lowest amongst those 65 and over (7.1\%). The experience of young people was therefore very mixed. Young people were among those groups hit hardest by job losses, and they were also the group whose firms were most likely to close. Yet, there was strong firm creation among this age category, too.
It is, however, important to put this acceleration in firm creation among young people into perspective. First, this growth in firm creation among young people was from a low baseline} (Figure \ref{fig:double_bars}, Table \ref{table:sic}). \remove{Second, n}\response{N}early half of new firms being created involved officers \remove{in middle-age (}\response{aged }35-49\remove{:} \response{(}40.4\%, Figure \ref{fig:double_bars}) and so the relative increase in younger entrepreneurs did very little to change the age distribution of firm control in the UK economy. Firm growth -- in absolute terms -- was largely driven by middle-aged men (through both the preservation of existing firms but also through new firms being created) even though there was strong growth in relative terms among young people. 

\begin{table}[H]
\begin{center}
    \caption{Summary statistics for firm outcomes by sector. The first four columns show pre-pandemic (August 2019 - February 2020) demographic data for officers split by sector: the proportion of officers that were women (\%F, column 1) and the age composition of officers (2-4). The last four columns show the change in the composition of companies by sector (March 2020 - June 2021): the change in the number of companies operating (5), opening (6), closing (7), and the share of companies operating in London (8). Values $>$1 denote an increase during Covid (relative to pre-Covid levels).}
    \begin{footnotesize}
    \begin{tabular}{p{.375\linewidth}ccccccccc}
        \toprule
        & \multicolumn{4}{c}{Pre-Covid} & & \multicolumn{4}{c}{Diff. in Pre-Covid and Covid}\\
        \cmidrule{2-5}\cmidrule{7-10}
        Sector & \%F & Age & \%$<$35 & \%$>$60 & & Firms & Open & Close & London\\ 
        \midrule
        A. Agriculture, Forestry and Fishing & .35 & 54.43 & .11 & .34 & & 1.13 & 1.34 & 0.98 & 1.07 \\
        B. Mining and Quarrying & .22 & 52.85 & .08 & .28 & & 1.15 & 1.01 & 0.85 & 1.19 \\
        C. Manufacturing & .27 & 51.55 & .13 & .24 & & 1.15 & 1.31 & 0.97 & 1.07 \\
        D. Electricity, Gas, Steam and Air Con. & .17 & 49.17 & .12 & .19 & & 1.37 & 1.36 & 0.93 & 0.83 \\
        E. Water Supply; Sewerage and Waste Mgmt & .23 & 50.95 & .13 & .23 & & 1.21 & 1.46 & 0.92 & 1.09 \\
        F. Construction & .23 & 48.82 & .16 & .19 & & 1.14 & 1.25 & 0.90 & 0.99 \\
        G. Wholesale \& Retail Trade; Vehicle Repair & .31 & 49.14 & .17 & .22 & & 1.22 & 1.73 & 0.95 & 1.07 \\
        H. Transportation and Storage & .22 & 44.85 & .25 & .15 & & 1.21 & 1.17 & 0.94 & 1.09 \\
        I. Accommodation \& Food Service Activities & .33 & 46.70 & .19 & .18 & & 1.23 & 1.14 & 0.88 & 1.04 \\
        J. Information and Communication & .25 & 46.14 & .20 & .15 & & 1.16 & 1.17 & 1.01 & 1.05 \\
        K. Financial and Insurance Activities & .23 & 50.85 & .10 & .21 & & 1.26 & 1.19 & 1.01 & 1.01 \\
        L. Real Estate Activities & .31 & 51.76 & .11 & .28 & & 1.22 & 1.29 & 1.01 & 1.02 \\
        M. Professional, Scientific and Technical & .33 & 50.42 & .12 & .21 & & 1.11 & 0.99 & 1.04 & 1.05 \\
        N. Administrative and Support Services & .33 & 46.82 & .19 & .16 & & 1.18 & 1.28 & 0.91 & 0.90 \\
        O. Public Administration and Defence & .38 & 49.65 & .15 & .20 & & 1.29 & 1.24 & 1.10 & 1.02 \\
        P. Education & .45 & 51.14 & .12 & .26 & & 1.28 & 1.15 & 0.99 & 1.04 \\
        Q. Human Health and Social Work & .48 & 49.45 & .12 & .22 & & 1.19 & 1.00 & 0.90 & 1.10 \\
        R. Arts, Entertainment and Recreation & .32 & 49.54 & .18 & .26 & & 1.17 & 1.10 & 0.89 & 0.98 \\
        S. Other Service Activities & .37 & 48.96 & .17 & .22 & & 1.20 & 1.09 & 0.99 & 1.07 \\
        T. Activities of Households as Employers & .44 & 52.40 & .11 & .32 & & 1.11 & 1.23 & 0.74 & 1.14 \\
        U. Activities of Extraterritorial Orgs & .30 & 51.16 & .14 & .28 & & 1.14 & 0.56 & 1.14 & 0.78 \\
        \bottomrule
    \end{tabular}
    \end{footnotesize}
    \label{table:sic}
    \end{center}
\end{table}

\subsection{Where did firm growth occur?}

\begin{figure}[H]
\begin{center}
\noindent\includegraphics[width=\textwidth]{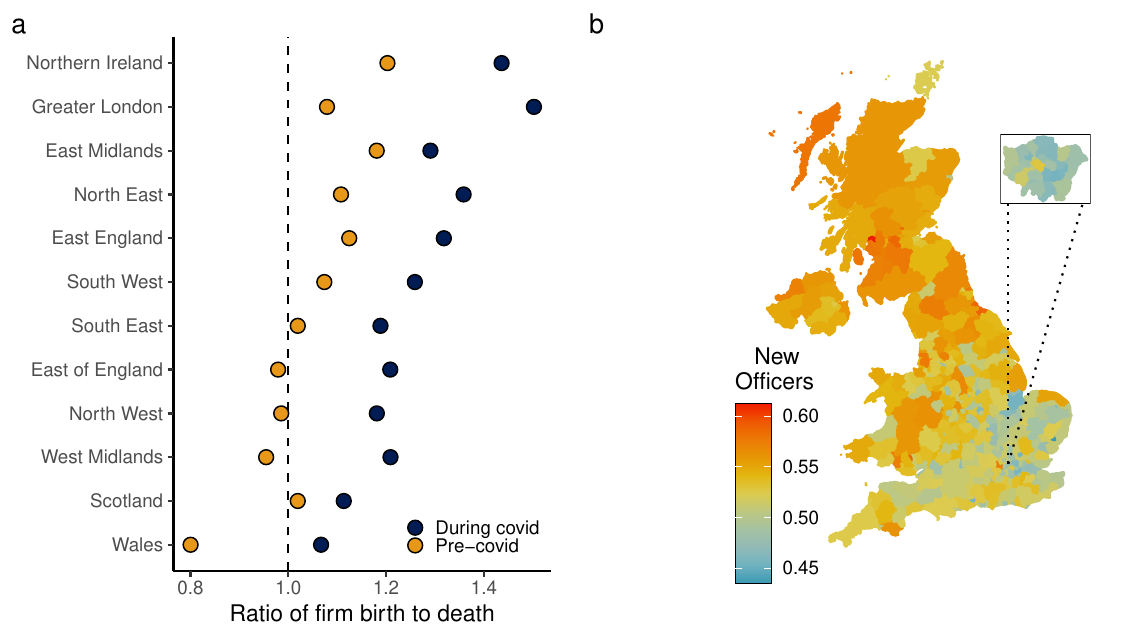}
    \caption{\response{Regional differences in firm outcomes. Panel a. shows the ratio of firm birth to death across the major regions of the UK both before and during the Covid pandemic. Panel b. shows the ratio of first-time officers to already-officers for all new companies created between March 2020 and June 2021. Darker colors denote a higher proportion of experienced officers, whereas yellower colors denote a higher proportion of first-time officers. Inset focuses on the Greater London area.}}
    \label{fig:map}
    \end{center}
\end{figure}

All regions of the UK experienced growth in officer numbers. This effect is observed across all age groups, driven by both an increase in firm openings and a reduction in firm closures (Figure \ref{fig:map}). The largest gains were seen in Greater London (39.2\% growth), Wales (33.3\%), and Northern Ireland (23.3\%). The growth in London-based companies can be seen across most aggregated Standard Industry Classifiers (SIC); for 15 out of 21 industries, the growth was higher in Greater London than the rest of the country (Table \ref{table:sic}). Growth was slowest in Scotland and the East Midlands (9.2\% and 9.3\% respectively, Figure \ref{fig:map}a.). Londoners were also more likely to officiate their first business during Covid (Figure~\ref{fig:map}b). This isn't the case in the rest of the country, where firm growth was mostly enjoyed by those who had prior experience of running firms.

\subsection{Is firm growth being driven by existing business owners?}

\begin{figure}[H]
    \begin{center}
    \includegraphics[width=\textwidth]{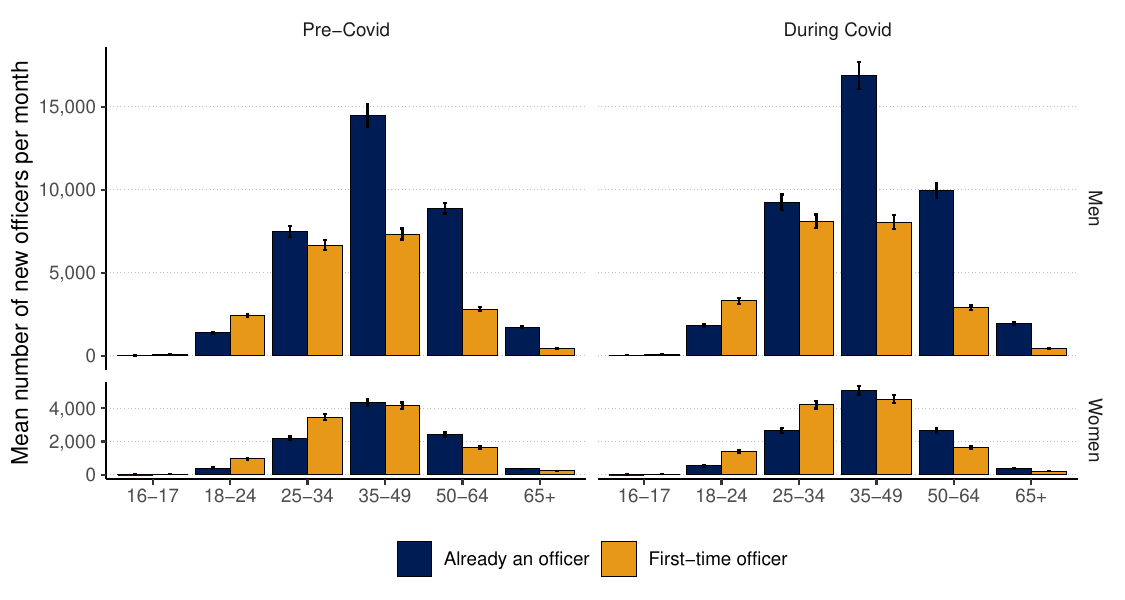}
        \caption{\response{Entrepreneurship among existing and first-time officers. Distribution of officers in new UK firms created in the months before Covid (August 2019--February 2020; `Pre-Covid') and during Covid (March 2020--June 2021; `During Covid'), categorized by whether they were `First-time officers' (individuals acting as an officer for the first time) or `Already an officer' (individuals who had previously served as an officer for at least one registered company). The data is further stratified by the officer’s age ($x$-axis), gender (vertical panel), and their prior officer status (color).}}
        \label{fig:new_officers}
    \end{center}
\end{figure}

\remove{In this final section of the results}\response{Next, we consider whether the rise in firm creation is being driven by first-time officers, which is what we might expect if the newly unemployed are transitioning to self-employment?}\remove{, we consider whether the simultaneous rise in firm creation and decline in employment are two sides of the same coin: that is, are the newly unemployed transitioning to self-employment, and in the process creating new firms?} This is not borne out by the data. Figure~\ref{fig:new_officers} shows that most new firms (59.4\%) set up in the UK were operated by people who already had an officer-ship in an existing firm. Not, \remove{for example}\response{as some speculated}, young entrepreneurs who may have recently been made unemployed. \remove{When a new firm is created, we can tell whether each officer in that new firm has previously been an officer in another firm. }\remove{We can also split this by gender to see what proportion of them are becoming officers for the first time. For men, the proportion of officers in a new firm who are becoming officers for the first time has been relatively steady at approximately approximately ~7\% Figure~\ref{fig:new_officers}.}

The trend over the Covid period is incredibly important because it reveals two divergent patterns. First, what is immediately noticeable is that \remove{right after Covid hit}\response{during Covid}, we see an increase in the proportion of new officers with experience running firms\response{, particularly among men aged 25-64} (Figure~\ref{fig:new_officers}). This very clearly suggests that the sharp increase in new firms created in the months following the start of the pandemic (Figure \response{\ref{fig:ch_time}}) was not created by the newly unemployed setting up firms. They were, instead, created by people who were already officers in existing firms.

\response{This is further reinforced by our analysis of firm creation probabilities across individuals with varying levels of prior firm ownership, presented in Table \ref{table:firm_ownership}. This shows that as the number of firms previously owned by an individual increases, so does their probability of establishing another firm during the pandemic. For instance, individuals who ran one firm on the eve of the pandemic (February 2020) had a 4.8\% probability of creating a new firm, whereas those with two prior firms had a 10\% probability. This probability continues to rise, with individuals owning ten or more firms exhibiting a 91.4\% probability of creating an additional firm in the 16 months following the onset of the pandemic (March  2020-June 2021). These patterns clearly indicate that existing firm owners disproportionately drove new firm creation during the pandemic.}

\remove{This is reinforced by the micro-level panel data which suggests that the proportion of the UK population that is self-employed has remained constant since 2019: firm growth cannot be attributed to an increase in newly unemployed people setting up a new firm.}

\subsection{\response{Industry-specific effects}}

\begin{figure}[!t]
    \begin{center}
        \includegraphics[width=\textwidth]{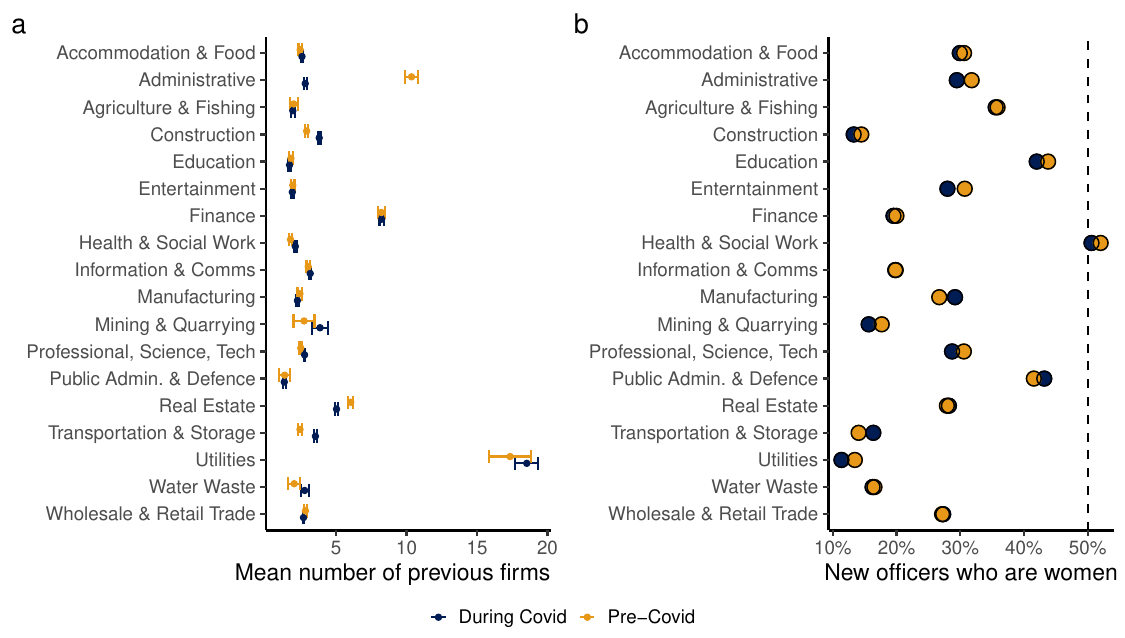}
    \end{center}
    \caption{\remove{Firm births, deaths ,and officerships across place and SIC. Sub-Figure a shows the ratio of firm birth to death across the major regions of the UK both before and during the Covid pandemic. Sub-Figure b shows the proportion of new officers who are women across major Standard Industry Classifier codes.}\response{Composition of new firms as a function of SIC and time period (August 2019--February 2020, `Pre-Covid'; March 2020--June 2021, `During Covid'). Panel a. shows The mean number of previous firms held by officers of new firms within a given industry (error bars denote 95\% confidence intervals). Number of firms within a given industry are winsorised at the 99.9th percentile. Panel b. shows the proportion of new officers that are women as a function of industry and time period, where the dashed line at 50\% denotes parity among men and women.}}\label{fig:sic}
\end{figure}

\response{Finally, in order to investigate the heterogeneity across industries and its potential effects on firm creation during the pandemic, we grouped new firms by SIC codes and compared firms opened before Covid (August 2019--February 2020) and after the onset of the pandemic (March 2020--June 2021). We measured the level of prior experience of individuals founding new firms by calculating the mean number of firms previously operated by officers of these firms, winsorising the data at the 99.9$^th$ percentile to mitigate the impact of outliers. Our analysis revealed substantive industry-level differences (Figure~\ref{fig:sic}b).
For example, at all time points, industries such as Utilities exhibited higher levels of prior entrepreneurial experience, whereas industries such as Education and Entertainment had relatively lower levels.

Furthermore, we observed significant heterogeneity in changes to this measure post-pandemic. For eleven industries, including Manufacturing and Information and Communication, the level of prior experience remained constant. However, industries like Construction, Health and Social Work, Professional Science and Technology, and Transport and Storage saw increases in the mean number of previous firms, suggesting an influx of more experienced individuals. Conversely, in Administrative, Wholesale and Retail Trade, and Real Estate industries, the level of prior entrepreneurial experience decreased, indicating a shift towards less experienced entrants. These findings underscore the importance of industry-specific factors in shaping the patterns of firm creation during the pandemic and highlight the differential barriers to entry and resource requirements across sectors.}

\response{Does the picture change when we look at whether industry-level effects are gendered? Not substantially. There remains substantial gender stratification across SIC groups in terms of the percentages of new officers who were women, both before and throughout the Covid period. Figure \ref{fig:sic}a, for example, shows that over half of the new officers in the field of Health and Social Work were female (a category which subsumes nursing and social care), while it was under 15\% in historically male dominated SIC areas such as Construction and Utilities. The proximity of the two dots for all major SIC groupings in Figure \ref{fig:sic}a show that the pandemic did not in any substantial way alter these gendered divides.} Firm death and creation over this period was very similar for both groups, and even the timing of the net growth in firm numbers is not radically different when stratified by gender.

\subsection{\response{Summary of results}}

\response{Putting these industry level differences to one side, what comes through strongly from our analysis is that}\remove{Together, these findings suggest that} most `excess firms' are run by people who already had a stake in existing firms, and not by an increase in aspiring entrepreneurs using Covid lock-downs to start new businesses. Second, there are important gender differences in firm creation over time. At a very basic level, among women, the proportion of officers in a new firm who are becoming officers for the first time is much higher ($\sim$50\%) than it is for men ($\sim$37\%). This is because women are, in general, much less likely to be an officer at all compared to men in the UK (Table \ref{table:sic}).\remove{ Crucially, this starts to change as the pandemic progresses. By mid-2020, the proportion of women officers in new firms who are becoming officers for the first time bounces back to the pre-pandemic level. But this starts to decline, suggesting at the very least that the pandemic has not fostered a generation of new women entrepreneurs.} Third, in terms of age, it is true that officers under the age of 25 in new firms were typically first-time officers, but these are a relatively small proportion of the people who are creating new firms. Most new firms were established by people aged 35 and older and in this category the officers of new firms were far more likely to already have experience of being an officer in one or more companies prior to the Covid recession (Figures \ref{fig:map}b-\ref{fig:map}c). It was existing business owners (those nascent `corporate elites') and not `Covid entrepreneurs' that were able to respond most rapidly to the pandemic.

\section{Discussion\remove{ and conclusion}}\label{sec:discussion}

If computational social scientists, economic sociologists, and policy\remove{ }makers want to understand wealth inequality, they need to attend to the dynamics of firm ownership. Using a computational approach, we examined who, if anyone, was able to capitalize on the economic crisis associated with Covid. \remove{Contrary to expectations, the pandemic led to outcomes that may have long-term effects on wealth inequality. }We \response{exploit}\remove{use} administrative data on the entire population of firms in the UK to offer new insights into how corporate elites responded \response{to} this pandemic-induced recession in terms of firm creation, paying particular attention to how this has been stratified by age, gender, and region. We focus on who became a `Covid entrepreneur'\remove{, contrasting this with micro- and macro-level data. Jointly considering the labor market impacts of the pandemic alongside firm-level impacts is critical for our understanding of the stratified effects of the pandemic on the economy} \citep{Adams2023}. Our results reinforce the perception that the group that witnessed the fewest economic losses in the UK were those in the middle of their working lives (35-49). Not only have they seen the smallest losses in employment and earnings, but they also saw `gains' in terms of firm ownership. The biggest losers were those under 30. Although they experienced an increase in firm creation, this was comparatively small relative to the economic losses felt by those under 30 in the labor market.

Covid-19 -- and the responses to it -- served to deepen the degree of economic insecurity faced by young people in the UK\remove{,} and \remove{therefore }may \response{also} have exacerbated \remove{generational }economic cleavages \response{between generations}\remove{ that are creating unequal cohort effects}. We \remove{also }show that this growth in firm creation has been led by those already holding corporate capital. This produces an increase in the concentration of corporate control\remove{,} \response{which is consistent} \remove{aligned }with theories of elite reproduction \citep{Useem1986}\remove{,} \response{and which }demonstrat\remove{ing}\response{es} how economic crises more generally can solidify the power of established elites\remove{, further entrenching class divisions}. This is not merely driven by the decline in firm deaths that would have occurred in the absence of policy changes (e.g., the moratorium on firms being removed from the register and the additional support offered by government to businesses). Nor is this driven by new firms solely created to take advantage of these new forms of government support (although further research is needed on how extensive this was). Instead, we document a nontrivial increase in the rise of firm creation among\remove{st} those who already had a significant stake in an existing firm\remove{, although it is yet to be seen how many of these `necessity-motivated' firms survive} \citep{Devece2016}.

\begin{table}[!t]
    \centering
    \caption{\response{Firm creation probabilities during the pandemic as a function of pre-existing firm ownership. Total number of individuals with varying levels of pre-existing firm ownership before the pandemic, the number of new firms they created during the pandemic, and their corresponding probabilities of firm creation. Probabilities are calculated as the ratio of new firms created to the pre-pandemic total for each category.}}
    \footnotesize
    \begin{tabular}{lrrr}
        \toprule
        N Pre-Existing Firms & Pre-Pandemic Total & Firms Created During Pandemic & Creation Prob. (\%) \\
        \midrule
        1 Firm  & 6,854,425 & 330,058 & 4.8 \\
        2 Firms & 1,425,875 & 142,590 & 10.0 \\
        3 Firms &   514,232 &  77,278 & 15.0 \\
        4 Firms &   238,446 &  47,044 & 19.7 \\
        5 Firms &   128,283 &  30,946 & 24.1 \\
        6 Firms &    76,088 &  21,677 & 28.5 \\
        7 Firms &    48,469 &  15,553 & 32.1 \\
        8 Firms &    33,017 &  12,028 & 36.4 \\
        9 Firms &    23,358 &   9,760 & 41.8 \\
        10+ Firms &  98,390 &  89,924 & 91.4 \\
        \bottomrule
    \end{tabular}
    \label{table:firm_ownership}
    \color{black}
\end{table}

\response{The growth in firm creation among the corporate elite has implications for the competing explanations we described in the introduction in part because the pandemic created a set of conditions which were very favorable to first-time entrepreneurs. In our view, if we cannot see a rise in the proportion of new firms being created by first-time officers during the pandemic then it is very unlikely that we would see a rise of this kind during normal recessions. In other words, \remove{that}\response{the} fact that corporate elites were \remove{able to }still able to capitalize on this crisis to a greater extent than new Covid entrepreneurs suggests that these same patterns would be observed in earlier recessions when the conditions were even less conducive to new entrants. The broader theoretical takeaway of this result is that corporate elites are able to capitalize on crises, even in conditions that should have undermined the power of existing corporate elites.}

Our analysis also suggests that the pandemic has likely increased regional inequalities\remove{, too} \citep{Dorling2022}. While firm growth occurred everywhere, London was the area with the largest amount of firm growth. This increased concentration of corporate control will almost certainly increase future wealth inequality in the UK because changes in wealth inequalities today are durable \citep{Beckert2022}. In addition, the impact of the pandemic on corporate elites may even undermine efforts to `level up' the UK economy because of the disproportionate degree of firm growth in and around London \citep{Fransham2023}.

Our empirical results contribute to our understanding of how corporate elites respond to economic uncertainty\remove{, and contributes to an evidence-base for future policy-makers during times of crisis}. While our data do not give us direct insight into exactly what specific entrepreneurs were thinking when deciding to either keep open a once failing business or to start a new firm, our data does provide two clues pertaining to why corporate elites are able to capitalize on crises \citep{Krippner2012}. The first concerns the role of corporate welfare \citep{Farnsworth2013}. The persistence of businesses that would have likely closed without government support suggests that corporate welfare played a key role in ensuring firms were able to keep operating. And yet, corporate welfare is not the only explanation for the persistence of these businesses. If it were purely a corporate welfare effect, then we would have expected to see these firms close once the payments from government stopped. But this is not what we see, and this is the second clue, which suggests that this was not the sole or even the main driver of the persistence of these businesses. Corporate welfare, then, potentially plays an important role in allowing elites to capitalize on these crises \citep{Farnsworth2013} but it is an inadequate explanation on its own.

Alongside government support, we also see evidence that corporate elites are situated in structurally advantaged positions that make it easier for them to have successful businesses. This resonates with capitalist class dynamics, highlighting how existing wealth and resources perpetuate the success of \remove{the capitalist class}\response{those with access to capital}, especially during times of crisis \citep{Wright1997}. Part of this is likely to be `know-how'. Setting up and running a business is substantially easier and more likely to be successful if you have been through the process before. On top of this, however, is the fact that wealth begets wealth \citep{DarityJr2018}. The sheer fact of being attached to a business already indicates that \response{a business owners }wealth is likely higher \remove{compared to}\response{than many} others in the population and, even where wealth is not higher than the average, it is probably still true that being an officer in an existing firm will make access to investment easier. \remove{We argue that the implications of our results have implications beyond the Covid pandemic. It is true that the pandemic-induced recession was different from previous recessions, but those differences seem to have potentially undermined the power existing corporate elites and create more opportunities for the redistribution of wealth through firm creation. In other words, that fact that corporate elites were able to still able to capitalize on this crisis to a greater extent than new Covid entrepreneurs suggests that these same patterns would be observed in earlier recessions when the conditions were even less conducive to new entrants.} Another implication of our results concerns the distribution of wealth \citep{Beckert2022}. \response{Many firms that likely would have closed in the absence of the pandemic}\remove{We find that many firms that would have closed in the absence of the pandemic} are still operating a year later, and there is no evidence of an acceleration in firms' deaths \response{in the months since}. Given the impact of firm ownership on wealth inequality, the continuation of these firms may have a small but nontrivial impact on the wealth distribution. \remove{In this respect, the}\response{The} forms of corporate welfare offered to these firms may have inadvertently accentuated wealth inequalities over the medium term. In addition, the fact that many of the\response{se} new firms \remove{that} have been created by \response{members of the corporate elite}\remove{those who already owned firms likewise }suggests that the pandemic will exacerbate wealth inequalities through more than just it's differential impact on savings. It is becoming increasingly clear that wealth has a profound impact on several different social outcomes, including social mobility \citep{Beckert2022}. \remove{The impact of the pandemic on firm growth and the labor market will, of course, be rooted in the internationally heterogeneous policy responses of specific governments. }Our findings\response{, then,} contribute to \remove{the}\response{our} understanding of how economic policies can exacerbate wealth inequalities \citep{Shane2003}, where pre-existing capital and social networks play a  pivotal role in determining who benefits from economic upheavals. More work is\response{, of course,} required to understand exactly how policy responses have mitigated the potential economic losses for different groups within society. In the UK, at least, we infer that the government policy accentuated existing inequalities, partly through the consolidation of corporate control.

\subsection{\response{Limitations}\label{sec:limitations}}

\response{There are a number of important limitations to our analysis. Perhaps the most significant is our definition of the corporate elite. We operationalize this concept in terms of the number of firms over which a person exercises significant control but there are other ways to define this group. For example, we could have defined the corporate elite as those who are persons of significant control in socially significant (e.g., high revenue firms). One reason we have not taken this approach is because we could not reliably measure business revenues at scale. This does leave some unanswered questions for our analysis. At the same time, however, we have demonstrated the robustness of our analysis to different definitions of the corporate elite in terms of board interlocks, finding a consistent story. Assuming that large businesses are also more likely to have more persons of significant control (e.g., through a board of directors) then our analysis is partially capturing these dynamics and may not diverge from this alternative approach in a significant way.} \response{Another major limitation of our population based approach is that we rely on approximate methods of gender inference, and approximate matching of entities into geographical polygons based on user-reported registration details. However, we have no reason to believe that this causes any systematic bias, other than the slight potential for more firms to register addresses in large metropolitan economic centers than operate therein.} \response{In addition, there are a number of potentially confounding variables that might be influencing our results that we have not been able to control for in our models. This is because the Companies House data does not give us information about the people who do not start new firms. In this respect, our analysis is, at one level, largely descriptive. We cannot parse all of the reasons that might lead corporate elites to open new firms. Despite this, we can rule out with a high degree of confidence the theory that newly unemployed people were driving the acceleration in firm creation simply because there was no acceleration in firm creation among first-time officers. Our analysis cannot confirm the causal impact of Covid on firm creation but what it does reveal is that this crisis did not undermine the capacity of corporate elites to capitalize on a crisis.}
\response{\subsection{Conclusion}\label{sec:conclusion}}

\response{Our results have significant implications for understanding wealth inequality, generational cleavages, and the resilience of corporate elites in times of crisis. We argue that the Covid pandemic represents a case study in how crises can entrench economic inequalities, rather than disrupt them. By exploring firm creation across age, region, and prior corporate experience, we find consistent evidence that established elites capitalized on the pandemic while other groups, particularly the young, faced heightened economic precarity. These findings underscore the need for policies that address the structural advantages of corporate elites and foster opportunities for less experienced entrepreneurs. Future research should examine the longevity and broader socio-economic impact of these newly created firms to understand their contribution to wealth inequality.}

\subsection*{Conflicts of interest}

On behalf of all authors, the corresponding author states that there is no conflict of interest.

\subsection*{Data availability}

All data are from publicly available sources. 
National estimates of corporate control by month (January 2019 to June 2021) are derived from Companies House's API (\url{https://developer.company-information.service.gov.uk}).
Household survey responses are derived from the `Understanding Society COVID-19' study (\url{https://www.understandingsociety.ac.uk/topic-page/covid-19/}).
National estimates of employment data by quarter are available from the Office of National Statistics (ONS series X01-X03, based on the Labour Force Survey; \url{https://www.ons.gov.uk/employmentandlabourmarket/peopleinwork/employmentandemployeetypes/datasets/labourforcesurveysinglemonthestimatesx01}).
Due to licensing agreements with Genderize and the large size of the raw data, we are unable to provide the raw data directly. However, aggregate data is available on GitHub (\url{https://www.github.com/nmuggleton/covid-inequality}). The raw data is stored on the University of Oxford's Advanced Research Computing cluster and can be made available upon \response{request}. 

\begingroup
\sloppy
\bibliography{refs}
\endgroup
\end{document}